\documentclass[reprint,amsmath,amssymb,aps,pra,showpacs]{revtex4-1}

\usepackage{graphicx}
\usepackage{dcolumn}
\usepackage{bm}
\usepackage[utf8]{inputenc}
\usepackage{amsfonts}
\usepackage{braket}
\usepackage{tabularx}
\usepackage[normalem]{ulem}

\begin{document}
\title{Influence of ion movement on the bound electron $g$-factor}

\author{Niklas Michel}\email{nmichel@mpi-hd.mpg.de}
\author{Jacek Zatorski}
\author{Christoph H. Keitel}

\affiliation{Max~Planck~Institute for Nuclear Physics, Saupfercheckweg 1, 69117 Heidelberg, Germany}

\date{\today}

 	
\begin{abstract}
In the relativistic description of atomic systems in external fields the total momentum and the external electric field couple to the angular momentum of the individual particles. Therefore, the motional state of an ion in a particle trap influences measurements of internal observables like energy levels or the $g$-factor. We calculate the resulting relativistic shift of the Larmor frequency and the corresponding $g$-factor correction for a bound electron in a hydrogen-like ion in the 1S state due to the ion moving in a Penning trap and show that it is negligible at the current precision of measurements. We also show that the analogous energy shift for measurements with an ion in the ground state of a Paul trap vanishes in leading order.
\end{abstract}

\maketitle

\section{Introduction}
Charged particle traps like the Penning- or Paul trap are devices which enable experiments on atomic structure theory with extraordinary precision. For example, the $g$-factor of an electron bound in hydrogen-like silicium has recently been measured with a $5\times 10^{-10}$ relative accuracy \cite{Sturm2011} and the atomic mass of the electron bound in hydrogen-like carbon has been derived from its $g$-factor with unprecedented relative precision of $3 \times 10^{-11}$ \cite{Jaceknature}. With a Paul trap the frequency of a mercury hyperfine transition was measured with a relative accuracy of $6\times 10^{-15}$ \cite{hfspaultrapnew}. Anticipating further improvements in the future, it is important to obtain theoretical estimates of all contributions to the measured observables.\\
Starting from a generalized Breit-Pauli-Hamiltonian \cite{Pachucki2004} that describes an atomic system in an external electromagnetic field including the leading relativistic corrections, the total momentum of the atom couples to the internal angular momenta of the constituents \cite{Pachucki2007,Pachucki2008}.\\
Since in experiments with trapped ions both the external field and the total momentum of the ion are nonvanishing, this has to be taken into account when the measurements involve states with different angular momenta. To the best of our knowledge this correction has not been evaluated for measurements of the bound electron $g$-factor in Penning traps and hyperfine splittings in Paul traps, which is calculated below.
\section{Coupling of angular momentum to external field and total momentum}
Due to leading relativistic corrections the angular momentum of an electron bound in an atomic system couples not only to an external magnetic field but also to an electric one, if the total momentum of the atom is nonvanishing. The relevant part of the effective Hamiltonian is taken from equation ($50$) in ref. \cite{Pachucki2007} and reads as
\begin{eqnarray}
\delta H:=&&
\sum_{a}\left(\frac{e_a}{2m_a}(\vec{l}_a+g_a \vec{s}_a)-\frac{Q}{2M}(\vec{l}_a+\vec{s}_a)\right)\nonumber\\
&&\cdot\,\frac{1}{2Mc^2}\left( \vec{\Pi} \times \vec{E} - \vec{E} \times \vec{\Pi} \right),
\label{eq:correction}
\end{eqnarray}
where $c$ is the speed of light, $a$ labels the different constituents of the atom, and $e_a$, $m_a$, $g_a$,  $\vec{l}_a$, and $\vec{s}_a$ are the corresponding charges, masses, $g$-factors, orbital angular momenta and spins, respectively. $Q$, $M$, $\vec{\Pi}=\vec{P}-Q\vec{A}$ are the total charge, mass and momentum of the atom, respectively. $\vec{E}$ is the external electric field and $\vec{A}$ the external vector potential.\\
That the operator from (\ref{eq:correction}) can influence $g$-factor measurements is proposed in \cite{Pachucki2008}. To obtain an estimate for this effect, we calculate the first order energy correction due to (\ref{eq:correction}) with the unperturbed state
\begin{equation}
\ket{\varphi_0}:=\ket{\text{int}}\otimes \ket{\text{ext}}
\end{equation}
being a direct product of the state $\ket{\text{int}}$ describing the internal degrees of freedom inside the atom, where the $\vec{l}_a$ and $\vec{s}_a$ operators act on, and the state $\ket{\text{ext}}$ describing the atom in the particle trap, where the $\vec{\Pi}$ and $\vec{E}$ operators act on. Thus, the first order energy correction due to (\ref{eq:correction}) reads
\begin{eqnarray}
\delta E =&& \bra{\varphi_0}\delta H\ket{\varphi_0}\nonumber \\
=&&\bra{\text{int}}\sum_{a}\left(\frac{e_a}{2m_a}(\vec{l}_a+g_a \vec{s}_a)-\frac{Q}{2M}(\vec{l}_a+\vec{s}_a)\right)\ket{\text{int}}\nonumber\\
&&\cdot\,\frac{1}{2Mc^2}\bra{\text{ext}}\left( \vec{\Pi} \times \vec{E} - \vec{E} \times \vec{\Pi} \right)\ket{\text{ext}}\label{eq:firstordercorrection}.
\end{eqnarray}
\section{Penning trap}
\begin{figure}[b]
\includegraphics[width=.375\textwidth]{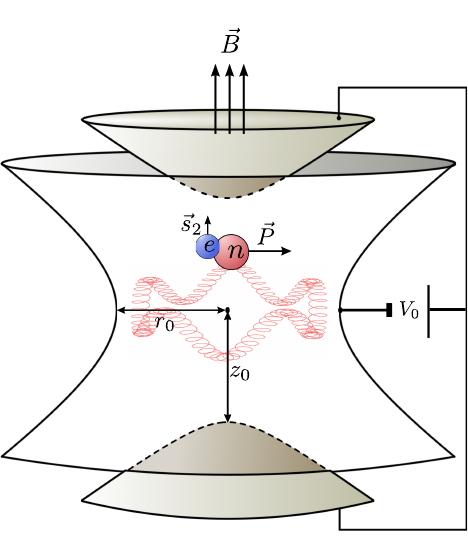}
\caption{\label{fig:penningtrap} Hydrogen-like ion with total momentum $\vec{P}$ in a Penning trap, where $e$ is the electron with spin $\vec{s}_2$, $n$ is the nucleus, and $\vec{B}$ the homogeneous magnetic field. $V_0$ is the voltage applied between the electrodes and the parameter $d^2=(z_0^2+r^2_0/2)$, where $r_0$ and $z_0$ are the distances between the center of the trap and the ring electrode and end cap electrode, respectively. The red line is a classical trajectory of the ion (color online).}
\end{figure}
In an ideal Penning trap \cite{chargedparticletrapsbook}, a charged particle is confined via a static homogeneous magnetic field
\begin{equation}
\vec{B}=B\,\vec{\text{e}}_z
\end{equation}
(let $B>0$) and an electric quadrupole potential
\begin{equation}
\Phi(\vec{r}) = V_0 \frac{2z^2-(x^2+y^2)}{2d^2}
\end{equation}
with the applied voltage $V_0$ and the parameter $d$ depending on the trap geometry. The Hamiltonian for an ion with charge $Q$ and mass $M$ is
\begin{equation}
H = \frac{1}{2M}(\vec{P}-Q\vec{A})^2 + Q \Phi.
\end{equation}
One possible choice for the vector potential $\vec{A}$ is
\begin{equation}
\vec{A}= \frac{1}{2}\vec{B}\times \vec{r} = \frac{B}{2}(-y\vec{\text{e}}_x+x\vec{\text{e}}_y).
\label{eq:afield}
\end{equation}
As a consequence, the Hamiltonian can be written as
\begin{eqnarray}
H=&&\frac{1}{2M}(p_x^2+p_y^2)+\frac{M\omega_1^2}{8}(x^2+y^2)-\frac{\omega_0}{2}L_z\nonumber\\
  &&+\frac{p_z^2}{2M}+\frac{M\omega_z^2}{2}z^2
\label{eq:harmoschamiltonian}
\end{eqnarray}
with $\omega_1^2 = \omega_0^2-2\omega_z^2$, $\omega_z^2 = 2QV_0 / Md^2$ and $\omega_0 = QB / M$. A hydrogen-like ion in a Penning trap with the parameters of the trapping potential is shown in Figure \ref{fig:penningtrap}. By choosing appropriate ladder operators, the Hamiltonian obtains the form of three uncoupled harmonic oscillators. The ladder operators are defined as
\begin{align}
a_z &=\frac{1}{\sqrt{2}z_1}\left(z+\frac{i}{M\omega_z}p_z\right),\\
a_{\pm} &=\frac{1}{2r_1}\left(x\pm i \frac{Q}{|Q|}y+\frac{2i}{M\omega_1}(p_x\pm i\frac{Q}{|Q|}p_y)\right),
\label{eq:ladder}
\end{align}
where $r_1=\sqrt{2\hbar / M\omega_1}$ and $z_1=\sqrt{2\hbar / M\omega_z}$.
The commutation relations are
\begin{equation}
[a_+,a_+^\dagger]=[a_-,a_-^\dagger]=[a_z,a_z^\dagger]=1,
\end{equation}
and 0 for all other combinations. Then the Hamiltonian reads
\begin{eqnarray}
H=&&\hbar\omega_+\left(a_+^\dagger a_++\frac{1}{2}\right)
 -\hbar\omega_-\left(a_-^\dagger a_-+\frac{1}{2}\right)\nonumber \\
 &&+\hbar\omega_z\left(a_z^\dagger a_z+\frac{1}{2}\right),
\end{eqnarray}
with $\omega_{\pm}= (|\omega_0| \pm \omega_1)/2$.\\
A complete orthogonal set of eigenstates is
\begin{equation}
\ket{nlk} = \frac{1}{\sqrt{n!l!k!}} (a_+^\dagger)^n (a_-^\dagger)^l (a_z^\dagger)^k \ket{0},
\label{eq:penningstate}
\end{equation}
with $\ket{0}$ being the groundstate of the ion in the trap. The corresponding energy eigenvalues are
\begin{equation}
E_{nlk} = \hbar\omega_+\left(n+\frac{1}{2}\right)
 -\hbar\omega_-\left(l+\frac{1}{2}\right)
 +\hbar\omega_z\left(k+\frac{1}{2}\right)
\end{equation}
and the angular momentum eigenvalues are
\begin{equation}
L_z = (l-n)\frac{|Q|}{Q}\hbar.
\end{equation}
To obtain the correction to the $g$-factor for a bound electron in the $1S$ state in hydrogen-like ions with a spinless nucleus the energy difference $\Delta E$ between a spin-up and spin-down state of the electron due to (\ref{eq:firstordercorrection}) has to be calculated (a=1 for nucleus, a=2 for electron). Since the spin $\vec{s}_1$ vanishes for a spinless nucleus, as well as the angular momenta $\vec{l}_1$ and $\vec{l}_2$ for an atomic system in the 1S state, only the electron spin $\vec{s}_2$ has a nonzero expectation value. For the spin-up state of the electron relative to the $\vec{B}$ field it is $\braket{s_{2,z}}=\hbar /2$ and for the spin-down state $-\hbar /2$ for the $z$-component and zero for the $x$- and $y$-component. Writing the state of the ion in the Penning trap $\ket{\text{ext}}$ as $\ket{nlk}$ via (\ref{eq:penningstate}), we obtain
\begin{eqnarray}
\Delta E=&&\left(\frac{e_2}{2m_2}g_2 -\frac{Q}{2M}\right)\hbar \nonumber \\ 
&&\cdot\,\frac{1}{2Mc^2}\Bra{nlk}\left( \vec{\Pi} \times \vec{E} - \vec{E} \times \vec{\Pi} \right)_z \Ket{nlk},
\label{eq:spinupdownenergycorrection}
\end{eqnarray}
\begin{table*}
\caption{\label{tab:penningresults}Results for the $g$-factor correction $\delta g$ compared to the experimental uncertainty $\Delta g_{exp}$ for different experiments}
\begin{ruledtabular}
\begin{tabular}{ccccccccc}
Ion & Ref. & $\omega_z$ [kHz]& $\omega_+$ [MHz] & $\omega_-$ [kHz] & $n$ & $l$ & $\Delta g_{exp}$&  $\delta g$ \\ \hline \\[-1.5ex]
$^{12}\rm{C}^{5+}$ & \cite{phdHaeffner,haffnercarbon} & $929 \times 2 \pi$ & $24 \times 2 \pi$ & $18 \times 2 \pi$ & $3.0 \times 10^7$ & $1.3 \times 10^8$ &$5 \times 10^{-9}$& $1.6 \times 10^{-12}$ \\
$^{16}\rm{O}^{7+}$ & \cite{phdVerdu,verduoxygen} & $925\times 2 \pi$ & $25\times 2 \pi$ & $17\times 2 \pi$ & $3.4 \times 10^3$ & $1.4 \times 10^6$ &$5 \times 10^{-9}$& $1.6 \times 10^{-16}$ \\
$^{28}\rm{Si}^{13+}$ & \cite{phdSturm,Sturm2011} & $705\times 2 \pi$ & $27\times 2 \pi$ & $9.3\times 2 \pi$ & $5.9 \times 10^4$ & $5.2 \times 10^4$ &$1 \times 10^{-9}$& $1.2 \times 10^{-15}$ \\ 
\end{tabular}
\end{ruledtabular}
\end{table*}
where $(\dots)_z$ denotes the $z$-component of a vector. The expectation value of $\vec{\Pi}$ and $\vec{E}$ can be obtained by writing them in terms of the ladder operators from (\ref{eq:ladder}).

The $g$-factor is obtained by measuring the Larmor frequency and the cyclotron frequency \cite{annphysgfactor}. The Larmor frequency corresponds to the energy difference of the spin-up and spin-down state relative to the $z$-component of the $\vec{B}$ field. For experiments with an ion in a Penning trap, here the coupling of the electron's spin to the total momentum of the ion has to be taken into account. Thus, the connection between Larmor frequency and $g$-factor \cite{annphysgfactor} with the correction from (\ref{eq:spinupdownenergycorrection}) becomes
\begin{equation}
h \nu_l = g \mu_b B + \Delta E,
\label{eq:larmor}
\end{equation}
where $\mu_b=|e_2|\hbar / (2m_2)$ is the Bohr magneton and the magnetic field $B$ can be eliminated with the cyclotron frequency via the Brown-Gabrielse invariance theorem \cite{geoniumtheory}
\begin{equation}
\omega_c =2\pi\,\nu_c= \sqrt{\omega_+^2 + \omega_-^2 + \omega_z^2} = \frac{|QB|}{M}.
\label{eq:invariancetheorem}
\end{equation}
If (\ref{eq:larmor}) is now solved for $g$, we arrive at
\begin{equation}
g=2\frac{\nu_l}{\nu_c}\frac{|Q|m_2}{M|e_2|}\underbrace{-\Delta E \frac{2}{h\nu_c}\frac{|Q|m_2}{M|e_2|}}_{=:\delta g}.
\end{equation}
Now we reintroduce and evaluate the expression $\Delta E$ from (\ref{eq:spinupdownenergycorrection}) to obtain
\begin{eqnarray}
\delta g =&& \frac{\hbar}{Mc^2}\frac{\omega_z^2}{\omega_c}
\left( g_2 - \frac{Q}{e_2}\frac{m_2}{M} \right)\nonumber
\\&&\times\left( n-l+ \frac{1}{\sqrt{1-2\frac{\omega_z^2}{\omega_c^2}}} (n+l+1) \right).
\end{eqnarray}
Neglecting above the higher-order correction of the charge to mass ratio of the ion devided by the one of the electron, employing the Dirac value $g_2=2$ and using the typical hierarchy $\omega_z\ll \omega_c$ \cite{geoniumtheory}, we obtain the approximate expression
\begin{equation}
\delta g \approx \frac{2\hbar}{Mc^2}\frac{\omega_z^2}{\omega_c}(2n+1).
\end{equation}
As a consequence, the dominant contribution to $\delta g$ is proportional to the quantum number $n$ and the inverse of the ion mass.
Numerical values of this correction for parameters from several experiments compared to the achieved experimental uncertainty $\Delta g_{exp}$ can be found in Table \ref{tab:penningresults}. The cyclotron frequency is determined with (\ref{eq:invariancetheorem}) and the quantum numbers $n$ and $l$ can be estimated with the frequencies and energies of the eigenmodes from
\begin{equation}
E_{n} = \hbar\omega_+\left(n+\frac{1}{2}\right)~~\text{and}~~
E_{l} = -\hbar\omega_-\left(l+\frac{1}{2}\right).
\end{equation}
The experimental accuracy still needs to become about four orders of magnitude higher for carbon and seven orders of magnitude higher for oxygen and silicium ions, so that the correction $\delta g$ becomes significant.
The different orders of magnitude of the effect for the different experiments can be explained by the different quantum numbers and therefore the energy of the ion in the trap. If the quantum numbers $n$ and $l$ are reduced in future experiments, $\delta g$ will decrease further.
\section{Paul trap}
For the analogous effect in a Paul trap we use the model of an ideal Paul trap \cite{chargedparticletrapsbook}, where an ion is confined with an oscillating electric potential
\begin{equation}
\Phi (\vec{r},t)=\frac{U_0+V_0\,\text{cos}(\Omega t)}{2d^2}(2z^2-x^2-y^2),
\label{eq:paulpotential}
\end{equation}
where $U_0$ and $V_0\,\text{cos}(\Omega t)$ are the dc- and ac-parts of the voltage, respectively, and the parameter $d$ depends on the geometry of the trap. A method to derive an effective time-independent potential from a rapidly oscillating potential quantum-mechanically is presented in \cite{pauleffectivepotential}. Applying this method to the Schrödinger equation with the potential $Q \Phi$, with $Q$ being the charge of the trapped ion, results in an effective Hamiltonian
\begin{equation}
H_{eff}=\frac{p_x^2+p_y^2+p_z^2}{2M}+\frac{M}{2}\omega^2(x^2+y^2)+\frac{M}{2}\omega_z^2 z^2
\label{eq:effectivehamiltonian}
\end{equation}
with
\begin{equation}
\omega^2=\frac{Q^2V_0^2}{2M^2\Omega^2 d^4}-\frac{QU_0}{M d^2}
\end{equation}
and
\begin{equation}
\omega_z^2=\frac{2Q^2V_0^2}{M^2\Omega^2 d^4}+\frac{2QU_0}{M d^2}.
\end{equation}\\
Hence, if the effective Hamiltonian is written as
\begin{equation}
H_{eff}=\vec{P}~^2/2M~+~Q \Phi_{eff},
\end{equation}
the resulting effective electric field is
\begin{equation}
\vec{E}_{eff}=-\vec{\nabla} \Phi_{eff} = -\frac{M\omega^2}{Q}(x\vec{\text{e}}_x+y\vec{\text{e}}_y)-\frac{M\omega_z^2}{Q}z\vec{\text{e}}_z.
\label{eq:effectiveEfield}
\end{equation}
The effective Hamiltonian (\ref{eq:effectivehamiltonian}) can be written in terms of the usual creation and annihilation operators as
\begin{eqnarray}
\label{eq:effectivehamiltonian2}
H_{eff}&&=\hbar \omega \left(a^{\dagger}_x a_x+a^{\dagger}_y a_y+1\right)+ \hbar \omega_z \left(a^{\dagger}_z a_z+\frac{1}{2}\right)\\
&&= \hbar \omega \left(a^{\dagger}_- a_-+a^{\dagger}_+ a_++1\right)+ \hbar \omega_z \left(a^{\dagger}_z a_z+\frac{1}{2}\right), \notag
\end{eqnarray}
where $a_{\pm}:=(1/\sqrt{2}) (a_x \pm i a_y)$ was used to define states
\begin{equation}
\ket{n_-n_+n_z} = \frac{1}{\sqrt{n_-!n_+!n_z!}} (a_-^\dagger)^{n_-} (a_+^\dagger)^{n_+} (a_z^\dagger)^{n_z} \ket{0}
\label{eq:paulstate}
\end{equation}
with eigenenergies
\begin{equation}
E_{n_-n_+n_z}=\hbar \omega (n_-+n_++1)+\hbar \omega_z \left(n_z+\frac{1}{2}\right)
\end{equation}
and eigenvalues of the $z$-component of the angular momentum
\begin{equation}
L_z = (n_--n_+)\hbar.
\end{equation}
Evaluating (\ref{eq:firstordercorrection}) for the Paul trap with the external state $\ket{\text{ext}}$ being $\ket{n_-n_+n_z}$ from (\ref{eq:paulstate}) and using the effective $\vec{E}$-field from (\ref{eq:effectiveEfield}), the result for the motional energy shift becomes
\begin{eqnarray}
\label{eq:paulenergycorrection}
&&\delta E =\notag \\
&&\bra{\text{int}}\sum_{a}\left(\frac{e_a}{2m_a}(l_{a,z}+g_a s_{a,z})-\frac{Q}{2M}(l_{a,z}+s_{a,z})\right)\ket{\text{int}}\notag \\
&&\times\frac{\hbar}{d^2Mc^2}\left( U_0 - \frac{Q V_0^2}{2M\Omega ^2 d^2} \right) (n_+-n_-).
\end{eqnarray}
Obviously, this correction vanishes for all states with $n_+=n_-$, in particular for the motional ground state of the ion in the trap where $n_+=n_-=0$. It has been shown, that the cooled ion can be placed mostly in the motional ground state \cite{paulgroundstate}. Nevertheless, to get an estimate for the size of the effect, we assume the internal part to contribute a factor of the size of the Bohr magneton for the $z$-component and that $n_+ - n_- = 1$ and use the trapping parameters from \cite{hfspaultrapold}. The resulting correction of a transition frequency in a $^{199}$Hg$^+$ ion would be of the order of $10^{-11}$ Hz compared to the experimental accuracy from \cite{hfspaultrapold} being of the order of $10^{-3}$ Hz.
\section{Conclusion}
The relativistic coupling of spin to the external electric field and the total momentum of an ion in a trap is investigated in order to estimate the leading corrections to the energy levels for an ion in a Penning- or Paul trap. For an ion cooled to the motional ground state in a Paul trap this correction vanishes in first order pertubation theory. In a Penning trap the resulting correction of the bound electron $g$-factor becomes important when the experimental accuracy is improved by about four orders of magnitude for an experiment with hydrogen-like carbon and seven orders of magnitude for hydrogen-like oxygen and silicium. In contrast to the situation with a Paul trap, it is important to notice, that the calculated correction $\delta g$ is non-vanishing even for the ground state of an ion in a Penning trap.
\bibliography{references}
\end{document}